\documentclass[12pt,a4j]{article}
\usepackage[dvips]{graphicx}
\usepackage{enumerate}
\usepackage{amsmath}

\begin{document}\title{Dispersionless Limits of Some Integrable  Equations}
\author{Zhaidary Myrzakulova\footnote{Email: zhrmyrzakulova@gmail.com} \, and  Ratbay Myrzakulov\footnote{Email: rmyrzakulov@gmail.com}\\
\textsl{Eurasian International Center for Theoretical Physics and} \\ { Department of General \& Theoretical Physics}, \\ Eurasian National University,
Astana, 010008, Kazakhstan
}
\date{}
\maketitle
\begin{abstract}
This is a write-up of the lectures  on dispersionless equations in 1+1, 2+1 and 3+1 dimensions presented by one of us (RM) at "Eurasian International Center for Theoretical Physics" (EICTP). We provide pedagogical introduction to the subject and summarize well-known results and some recent developments in theory of integrable dispersionless equations.  Almost all results presented are available in the literatures. We just add some new results related  to  dispersionless limits of integrable magnetic  equations.  On the contrary, some of the basic tools of the integrable dispersionless  systems  and related theoretical techniques are not available in a pedagogical format. For this  reason we think that  it seems worthwhile to present basics of theory of dispersionless equations  here for the benefit of  beginner  students. We begin with a detailed exposition of the well-known dispersionless systems like dKdVE, dNLSE, dKPE, dDSE and other classical soliton equations.  We present in detail some new dispersionless systems. Next we develop the dispersionless limits of known magnetic equations. Lastly,  we discuss in full detail the Lax representation  formulation of  some presented dispersionless equations.  On the basis of the material presented here one can proceed smoothly to read the recent developments in this field of integrable dispersionless equations and related topics.

\end{abstract}
\newpage


\section{Introduction} 

Integrable nonlinear differential  equations play an important   role in both modern physics and mathematics (see, e.g. \cite{Zakharov}-\cite{1902.06975} and references therein). One of important subclass of integrable equations is the so-called dispersionless equations. Dispersionless equations in 1+1, 2+1 and 3+1 dimensions are  quasilinear systems of the following forms
\begin{eqnarray}
{\bf u}_{t} + A({\bf u}){\bf u}_{x} &=& f_{1},  \label{1}\\
{\bf u}_{t} + A({\bf u}){\bf u}_{x}+ B({\bf u}){\bf u}_{y} &=& f_{2},  \label{2}\\
{\bf u}_{t} + A({\bf u}){\bf u}_{x}+ B({\bf u}){\bf u}_{y}+ C({\bf u}){\bf u}_{z} &=& f_{3},  \label{3}
\end{eqnarray}
respectively. Here  $t,x,y,z$ are independent variables, ${\bf u}=(u_{1},u_{2}, ..., u_{m})$ is an $m$-component vector and $A({\bf u}), B({\bf u}), C({\bf u})$ are $m\times m$ matrices, $f_{j}(x,y,z,t, u_{k})$ are some real functions. 

This paper is a notes  of the lectures on integrable dispersionless equations presented by one of us (RM) at "Eurasian International Center for Theoretical Physics" (EICTP). We provide pedagogical introduction to the subject and summarizes some known important results and many recent developments in theory of integrable dispersionless equations.  Almost all results presented are available in the literatures. We just add some new results related to  dispersionless limits of known fundamental integrable magnetic  equations.  On the contrary, some of the basic tools of the integrable dispersionless  systems  and related theoretical techniques are not available in a pedagogical format, and it seems worthwhile to present them here for the benefit of   students. We begin with a detailed exposition of the well-known dispersionless systems like dKdVE, dNLSE, dKPE, dDSE and other classical integrable  equations.  We present in detail some new dispersionless systems. Next we develop the dispersionless limits of known integrable magnetic equations. Lastly, we discuss in full detail the Lax representation  formulation of the dispersionless equations.  On the basis of the material presented here one can proceed smoothly to read the recent developments in this field of integrable dispersionless equations and related topics.

The structure of the paper is as follows. In Section 2 we present  the
dispersionless limits of some integrable equations in 1+1 dimensions.  Then, in Section 3,  we give basic  informations on the Benney equation. Integrable dispersionless systems in 2+1 dimensions are  considered in Section 4.  We  discuss briefly (3+1)-dimensional integrable systems in Section 5.   Finally, in Section 6,  we construct the dispersionless limits of some integrable magnetic equations. Section 7 is devoted to the conclusion remarks and comments.

\section{Dispersionless equations in 1+1 dimensions}
\subsection{Dispersionless KdV equation}
Consider the KdV equation (KdVE)
\begin{eqnarray}
u_{t}+1.5 uu_{x}+u_{xxx}=0.  \label{4}
\end{eqnarray}
Its dispersionless limit is
\begin{eqnarray}
u_{t}+1.5 uu_{x}=0.  \label{5}
\end{eqnarray}
The dKdV equation is also called the Riemann equation. The Lax representation of the dKdV equation looks like
\begin{eqnarray}
L_{t}=\{M,L\}, \label{6}
\end{eqnarray}
where 
\begin{eqnarray}
 L&=&p^{2}+u, \label{7}\\
M&=&p^{3}+1.5up. \label{8}
\end{eqnarray}
In (\ref{6}), the Poisson bracket has the form
\begin{eqnarray}
 \{A(x,p),B(x,p)\}=\frac{\partial A}{\partial p}\frac{\partial B}{\partial x}-\frac{\partial B}{\partial p}\frac{\partial A}{\partial x}. \label{9}
\end{eqnarray}
\subsection{Dispersionless NLSE}
Consider the  $\epsilon$ - dependent NLSE
\begin{eqnarray}
i\epsilon q_{t} + \epsilon^{2}q_{xx} + 2\delta|q|^{2}q = 0,  \label{10}
\end{eqnarray}
where $\delta = \pm 1$.  We can rewrite this equation using   the Madelung transformation
\begin{eqnarray}
q =\sqrt{u}e^{i\frac{\partial^{-1}_{x}(v)}{\epsilon}}=\sqrt{u}e^{\frac{iS}{\epsilon}}, \label{11}
\end{eqnarray}
where $u(x, t)$ is  the density  and $v(x, t)$ is the flow velocity. Then in the limit $\epsilon\rightarrow 0$, the NLSE can be cast into a hydrodynamiclike form \cite{Zakharov}:
\begin{eqnarray}
 u_{t} + 2(uv)_{x}&=&0, \label{12}\\
v_{t} + 2vv_{x} - 2\delta u_{x}&=& 0. \label{13}
\end{eqnarray}
It is the dNLSE or the Benney equation. Its Lax representation has the form
\begin{eqnarray}
 L_{t} = \{M, L\}, \label{14}
\end{eqnarray}
where
\begin{eqnarray}
L&=&p+u+\frac{v}{p}, \label{15}\\
M&=&p^{2}+2up.  \label{16}
\end{eqnarray}
The related Lax pair is given by ($E=const$):
\begin{eqnarray}
E&=&R_{x}+u+vR^{-1}_{x}, \label{17}\\
R_{t}&=&R_{x}^{2}+2uR_{x}.\label{18}
\end{eqnarray}
Note that the dNLSE  can be written in a more symmetric form. To do that we introduce  the
Riemann invariants \cite{1902.06975}:
\begin{equation}
\lambda^{\pm}=\frac{u}{2}\pm\sqrt{\rho}. \label{19}
\end{equation}
The Riemann invariants obey the following equations
\begin{equation}
\partial_t\lambda^\pm +v_\pm(\lambda^-,\lambda^+)\, \label{20}
\partial_x\lambda^\pm=0, 
\end{equation}
where
\begin{equation}
 v_\pm(\lambda^-,\lambda^+)
  =\tfrac{1}{2}(3\lambda^\pm+\lambda^\mp)=u\pm\sqrt{\rho}. \label{21}
\end{equation}
We can  linearize this equations  by means of the hodograph
transform (see, e.g., Ref.~\cite{Kamchatnov}). We have 
\begin{equation}
\partial_\pm x -v_\mp \partial_\pm t=0,  \label{22}
\end{equation}
where $\partial_{\pm} \equiv \partial/ \partial \lambda^{\pm}$.
These equations can be rewritten as
\begin{eqnarray}
\frac{\partial_-  W_+ }{ W_+ -  W_-} &=& \frac{\partial_- v_+}{v_+-v_-}, \\   \label{23}
\frac{\partial_+  W_- }{ W_+ -  W_-} &=&\frac{\partial_+ v_-}{v_+-v_-},  \label{24}
\label{Tsarev}
\end{eqnarray}
where 
\begin{equation}
 W_\pm(\lambda^-,\lambda^+)=x - v_\pm(\lambda^-,\lambda^+)t.\label{25}
\end{equation}
It is the Tsarev equations \cite{Tsarev}. From Tsarev equations (\ref{23})-(\ref{24}) follows that
\begin{equation}
\partial_-
W_+ = \partial_+ W_-  \label{26}
\end{equation}
which  gives
\begin{equation}
  W_\pm = \partial_\pm \chi. \label{27}
\end{equation}
Here  $\chi(\lambda^-,\lambda^+)$ is some function (potential). In terms of $\chi$, the Tsarev equation takes the form
\begin{equation}
\frac{\partial^2 \chi}{\partial \lambda^+ \partial\lambda^-} -
\frac{1}{2\,(\lambda^+ - \lambda^-)}
\left( \frac{\partial \chi}{\partial \lambda^+} -
\frac{\partial \chi}{\partial \lambda^-} \right) = 0 \label{28}
\end{equation}
or
\begin{equation}
\frac{\partial^2 \chi}{\partial \lambda^+\partial \lambda^-}
+ a(\lambda^-,\lambda^+) \frac{\partial \chi}{\partial \lambda^+} +
b(\lambda^-,\lambda^+) \frac{\partial \chi}{\partial \lambda^-} = 0, \label{29}
\end{equation}
where 
\begin{equation}
a(\lambda^-,\lambda^+) = - b(\lambda^-,\lambda^+) =
- \frac{1}{2\,(\lambda^+ - \lambda^-)}. \label{30}
\end{equation}
It is nothing but  the Euler-Poisson equation \cite{1902.06975}. 

\subsection{Dispersionless Kaup-Newell equation}
The Kaup-Newell equation reads as
\begin{eqnarray}
iq_{t}+q_{xx}+i\delta(|q|^{2}q)_{x}=0, \label{31}
\end{eqnarray}
Its dispersionless limit is
given by
\begin{eqnarray}
u_{t}+2(uv)_{x}+\frac{3}{2}\delta (u^{2})_{x}&=&0, \label{32}\\
v_{t}+(v^{2}+\delta uv)_{x}&=&0. \label{33}
\end{eqnarray}
\subsection{Dispersionless  Chen-Lee-Liu equation}
The Chen-Lee-Liu equation is given by
\begin{eqnarray}
iq_{t}+q_{xx}+i\delta|q|^{2}q_{x}=0. \label{34}
\end{eqnarray}
Its dispersionless limit reads as 
\begin{eqnarray}
 u_{t}+2(uv)_{x}+\frac{1}{2}\delta (u^{2})_{x}&=&0, \label{35}\\
v_{t}+(v^{2}+\delta uv)_{x}&=&0. \label{36}
\end{eqnarray}
\subsection{Dispersionless  Yajima-Oikawa-Ma equation}
The Yajima-Oikawa-Ma equation looks like:
\begin{eqnarray}
iq_{t}+q_{xx}-wq&=&0, \label{37}\\
w_{t}+\sigma w_{x}+2\delta(|q|^{2})_{x}&=&0, \label{38}
\end{eqnarray}
where $\sigma=0/1, \quad \delta=\pm 1$. Its dispersionless limit is given by
\begin{eqnarray}
 u_{t}+2(uv)_{x}&=&0, \label{39}\\
v_{t}+(v^{2}+w)_{x}&=&0, \label{40}\\
w_{t}+\sigma w_{x}+2\delta u_{x}&=&0. \label{41}
\end{eqnarray}

\section{Benney equations}

The Benney equation (BE) has the following form
\begin{equation}
 A_{k,t} = A_{k+1,x} + k A_{k-1}
A_{0,x}.\label{42} 
\end{equation}
The Benney equation is integrable by the Lax representation
\begin{equation}
L_{t} = - {1\over 2}
\left\{(L^{2})_{+},L\right\} \label{43}
\end{equation}
where 
\begin{equation}
L = p + \sum_{k=0}^{\infty} A_{k} p^{-(k+1)}.\label{44}
\end{equation}
The BE admits some important reductions. For example, if we take 
\begin{equation}
A_{k} = u^{k} v, \label{45}
\end{equation}
then the BE reduces to the dNLSE
\begin{equation}
v_{t} = (uv)_{x},\quad  u_{t}= uu_{x} + v_{x}. \label{46}
\end{equation}
The BE  arises as the consistency condition of the following equations
\begin{eqnarray}
 \lambda &=& p + \sum_{k=0}^\infty A^{k}p^{-(k+1)}, \\ \label{47}
 p_{t_2}&=&- p p_x - A^0_x, \\ \label{48}
p_{t_3} &=&- p^2 p_x - (p A^0+A^1)_x. \label{49}
\end{eqnarray}
Hence, for example,  if we take $ u = A^0$ and $y=t_2, \quad t= t_3$,  then we get the dKPE. Note that the above presented reductions are described by the Gibbons-Tsarev equation.

\section{Dispersionless equations in 2+1 dimensions}

\subsection{Dispersionless Kadomtsev-Petviashvili equation}
Consider the Kadomtsev-Petviashvili equation (KPE) which reads as
\begin{eqnarray}
(u_{t}+uu_{x}+\sigma u_{xxx})_{x}+\delta u_{yy}=0. \label{50}
\end{eqnarray}
It is well-known that the dispersionless Kadomtsev-Petviashvili equation (dKPE) looks like 
\begin{eqnarray}
(u_{t}+uu_{x})_{x}+u_{yy}=0. \label{51}
\end{eqnarray}
Consider the following set of equations
\begin{eqnarray}
L_{1}\phi&=&0, \label{52}\\
L_{2}\phi&=&0, \label{53}
\end{eqnarray}
where $L_{j}$ are the one-parameter families of vector fields of the forms
\begin{eqnarray}
L_{1}&=&\partial_{y}+\lambda\partial_{x}-u_{x}\partial_{\lambda}=
\partial_{y}+\{H_{2},\cdot\}_{(\lambda,x)} \label{54}\\
L_{2}&=&\partial_{t}+(\lambda^{2}+u)\partial_{x}-(\lambda u_{x}-u_{y})\partial_{\lambda}=\partial_{t}+\{H_{3}.\cdot\}_{(\lambda,x)}, \label{55}
\end{eqnarray}
Here the Hamiltonians are given by
\begin{eqnarray}
H_{2}&=&\frac{1}{2}\lambda^{2}+u, \label{56}\\
H_{3}&=&\frac{1}{3}\lambda^{3}+\lambda u-\partial_{x}^{-1}(u_{y}).\label{57}
\end{eqnarray}
Then the compatibility condition of the system (\ref{9})-(\ref{9}) 
\begin{eqnarray}
[L_{1}, L_{2}]=0 \label{58}
\end{eqnarray}
gives the dKPE (\ref{9}). Finally we note that the dKPE can be also presented as the compatibility condition for the following quasilinear 
system \cite{1612.02753}
\begin{eqnarray}
\varphi_{x}&=&(\varphi^{2}-u)\varphi_{t}-u_{t}\varphi -u_{y}, \label{59}\\
\varphi_{y}&=&\varphi\varphi_{t}-u_{t}.\label{60}
\end{eqnarray}
In fact, the compatibility condition of these equations $\varphi_{xt}=\varphi_{tx}$ gives
\begin{eqnarray}
u_{tx}+(uu_{t})_{t}-u_{yy}=0\label{61}
\end{eqnarray}
that coinside with  the dKPE (\ref{9}) with some scale transformations.

\subsection{Dispersionless Davey-Stewartson  equation}
Our next example is the Davey-Stewartson  equation (DSE) with  the following form 
\begin{eqnarray}
iq_{t}+\frac{1}{2}(q_{xx}+\sigma^{2}q_{yy})+\delta\phi q&=&0, \label{62}\\
\sigma^{2}\phi_{yy}-\phi_{xx}+(|q|^{2})_{xx}+\sigma^{2}(|q|^{2})_{yy}&=&0 \label{63}
\end{eqnarray}
or
\begin{eqnarray}
iq_{t}+q_{zz}+q_{\bar{z}\bar{z}}+\delta\phi q&=&0, \label{64}\\
\phi_{z\bar{z}}-0.5[(|q|^{2})_{zz}+(|q|^{2})_{\bar{z}\bar{z}}]&=&0, \label{65}
\end{eqnarray}
where $q(x,y,t)$ is a complex (wave-amplitude) field,  $\phi(x,y,t)$ is a real (mean-flow) field and
\begin{eqnarray}
z=x+\sigma y, \quad \bar{z}=x-\sigma y. \label{66}
\end{eqnarray}
 Consider the following transformation
\begin{eqnarray}
q=\sqrt{u}e^{\frac{i}{\epsilon}S}. \label{67}
\end{eqnarray}
Then the dispersionless Davey-Stewartson  equation (dDSE) reads as \cite{Yi1}-\cite{Yi2}:
\begin{eqnarray}
u_{t}+(uS_{x})_{x}+(uS_{y})_{y}&=&0, \label{68}\\
S_{t}+0.5(S_{x}^{2}+\sigma^{2}S_{y}^{2})-\delta\phi&=&0, \label{69}\\
\sigma^{2}\phi_{yy}-\phi_{xx}+u_{xx}+\sigma^{2}u_{yy}&=&0, \label{70}
\end{eqnarray}
or
\begin{eqnarray}
u_{t}+2[(uS_{z})_{z}+(uS_{\bar{z}})_{\bar{z}}]&=&0, \label{71}\\
S_{t}+S_{z}^{2}+S_{\bar{z}}^{2}-\delta\phi&=&0, \label{72}\\
\phi_{z\bar{z}}-0.5(u_{zz}+u_{\bar{z}\bar{z}})&=&0. \label{73}
\end{eqnarray}
Consider the following one-parameter Hamiltonian vector fields Lax pair
\begin{eqnarray}
L_{1}&=&\partial_{\bar{z}}-\{H_{1}, \cdot\}_{\lambda,z},
 \label{74}\\
L_{2}&=&\partial_{t}-\{H_{2}, \cdot\}_{\lambda,z}, \label{75}
\end{eqnarray}
where
\begin{eqnarray}
H_{1}&=&S_{\bar{z}}+\frac{\delta}{4}\frac{u}{\lambda},
 \label{76}\\
H_{2}&=&\lambda^{2}-2S_{z}\lambda\lambda+\frac{\delta}{2}W-S_{
\bar{z}}^{2}
-\frac{\delta}{2}\frac{uS_{\bar{z}}}{\lambda}-
\frac{1}{16}\frac{u^{2}}{\lambda^{2}}. \label{77}
\end{eqnarray}
The commutation condition
\begin{eqnarray}
[L_{1}, L_{2}]=0 \label{78}
\end{eqnarray}
gives the dDSE (\ref{71})-(\ref{72}).

\subsection{ Dispersionless NLSE in 2+1 dimensions}
Now we consider the following (2+1)-dimensional NLSE
\begin{eqnarray}
iq_t+q_{xy}-w q&=&0, \label{79}\\
w_x+2\delta\left(\left|q\right|^2\right)_y&=&0.  \label{80}
\end{eqnarray}
If $y=x$,  then Eqs.(\ref{79})-(\ref{80})  reduces to the (1+1)-dimensional NLSE:
\begin{eqnarray}
iq_{t}+q_{xx}+2\delta|q|^2q=0. \label{81}
\end{eqnarray}
Returning  to the Madelung transformation (\ref{11}), we obtain the following  dispersionless limit of the (2+1)-dimensional NLSE (\ref{79})-(\ref{80}):
\begin{equation}
\begin{cases}
S_{t}+S_{x}S_{y}+\omega=0,\\ 
u_{t}+2S_{xy}u+S_{x}u_{y}+S_{y}u_{x}=0,\\
\omega_x+2\delta u_{y}=0.\label{82}
\end{cases}
\end{equation}
In terms of $v=S_{x}\rightarrow S=\partial_{x}^{-1}v$, these equations take the form
\begin{equation}
\begin{cases}
u_{t}+2v_{y}u+vu_{y}+u_{x}\partial_{x}^{-1}v_{y}=0,\\
v_{t}+(v\partial_{x}^{-1}v_{y})_{x}+\omega_x=0,\\
\omega_x+2\delta u_{y}=0, \label{83}
\end{cases}
\end{equation}
or
\begin{equation}
\begin{cases}
u_{t}+2v_{y}u+vu_{y}+u_{x}z=0,\\
v_{t}+(vz)_{x}+\omega_x=0,\\
\omega_x+2\delta u_{y}=0,\\
z_x=v_y, \label{84}
\end{cases}
\end{equation}
where  $\partial_{x}^{-1}v_{y}=z, z_{x}=v_{y}$. If $y=x$,  then  $z=v, \omega=-2\delta u$, and  we obtain the (1+1)-dimensional dNLSE
\begin{equation}
\begin{cases}
u_{t}+2(uv)_x=0,\\
v_{t}+(v^2)_{x}-2\delta u=0. \label{85}
\end{cases}
\end{equation}

\section{Dispersionless equations in 3+1 dimensions}
Integrable nonlinear  partial differential equations in 3+1 dimensions play very important role in modern physics and mathematics. There are many examples of such integrable systems in literature (see, e.g.  \cite{SA1}-\cite{SA3} and references therein). The particular but important interest are  integrable dispersionless equations. 
In spite of this,  and avoiding technical difficulties, we decided to limit ourselves here only to bring the  in literatures on this subject.

\section{Dispersionless  magnetic equations}

In this section, we find the dispersionless limits of some integrable magnetic equations \cite{z1}-\cite{Gerdjikov}. We consider two examples of such magnetic equations: the Heisenberg ferromagnet equation (HFE) and the Landau-Lifshitz equation (LLE). The HFE has the form
\begin{eqnarray}
{\bf A}_{t}={\bf A}\wedge{\bf A}_{xx}, \label{86}
\end{eqnarray}
where ${\bf A}=(A_{1}, A_{2}, A_{3}) 
$ is a spin (magnetic) vector and ${\bf A}^{2}=1$.

The LLE reads as
\begin{eqnarray}
 {\bf A}_{t}={\bf A}\wedge{\bf A}_{xx}+{\bf A}\wedge J{\bf A}, \label{87}
\end{eqnarray}
where $J=diag(j_{1},j_{2},j_{3})$ is an arbitrary constant diagonal matrix.

To best of our knowledge, the dispersionless limits of integrable magnetic equations (including HFE and LLE) were unknown in literatures (see, e.g. Refs. \cite{z1}-\cite{z2}).
\subsection{M-XII equation }
The M-XII equation  has the form
\begin{eqnarray}
 u_{t}+ 2(uv)_{x}- \frac{4uvu_{x}}{1+u}&=&0, \label{88}\\
 v_{t}+2vv_{x}- \left(\frac{2uv^{2}}{1+u}\right)_{x}&=&0 \label{89}
\end{eqnarray}
or
\begin{eqnarray}
 u_{t}+ 2(uS_{x})_{x}- \frac{4uu_{x}S_{x}}{1+u}&=&0, \label{90}\\
 S_{t}+S_{x}^{2}- \frac{2uS_{x}^{2}}{1+u}&=&0, \label{91}
\end{eqnarray}
where $v=S_{x}$. Let us derive this equation. Consider the HFE
\begin{eqnarray}
{\bf A}_{t}={\bf A}\wedge{\bf A}_{xx}, \label{92}
\end{eqnarray}
where ${\bf A}=(A_{1}, A_{2}, A_{3}) 
$ is a spin (magnetic) vector and ${\bf A}^{2}=1$. The stereographic projection of the magnetic vector ${\bf A}$ is given by
\begin{eqnarray}
A^{+} =A_{1}+iA_{2}=\frac{2\omega}{1+ |\omega|^{2}}, \quad A_{3}=\frac{1-|\omega|^{2}}{1+ |\omega|^{2}}, \label{93}
\end{eqnarray}
or
\begin{eqnarray}
\omega=\frac{A^{+}}{1+A_{3}}.\label{94}
\end{eqnarray}
Then the HFE (\ref{11}) takes the form
\begin{eqnarray}
i\omega_{t}+ \omega_{xx} -
\frac{2\bar{\omega}\omega^{2}_{x}}{1+ |\omega|^{2}}= 0.\label{95}
\end{eqnarray}
Let us  rewrite this equation as
\begin{eqnarray}
i\epsilon\omega_{t}+ \epsilon^{2}\omega_{xx} -
\epsilon^{2}\frac{2\bar{\omega}\omega^{2}_{x}}{1+ |\omega|^{2}}= 0. \label{96}
\end{eqnarray}
Consider the transformation
\begin{eqnarray}
\omega =\sqrt{u}e^{i\frac{\partial^{-1}_{x}(v)}{\epsilon}}. \label{97}
\end{eqnarray}
Then the functions $u$ and $v$ obey the M-XII equation (\ref{88})-(\ref{89}). We can rewrite the M-XII  equation (\ref{88})-(\ref{89}) as
\begin{eqnarray}
 w_{t}+ 2(vw-v)_{x}- \frac{4(w-1)vw_{x}}{w}&=&0, \label{98}\\
 v_{t}+2vv_{x}- \left(\frac{2(w-1)v^{2}}{w}\right)_{x}&=&0 \label{99}
\end{eqnarray}
or
\begin{eqnarray}
 w_{t}+ 2(vw)_{x}-2v_{x}- 4vw_{x}+\frac{4vw_{x}}{w}&=&0, \label{100}\\
 v_{t}- 2vv_{x}+\left(\frac{2v^{2}}{w}\right)_{x}&=&0, \label{101}
\end{eqnarray}
where $w=1+u$.
Thus we have proved that M-XII equation (\ref{88})-(\ref{89}) is the dispersionless limit of the HFE (\ref{92}). It is the two-component dispersionless equation \cite{Brunelli1}.

\subsection{M-LXXIV equation}
Let us now we consider  the following M-LXXIV  equation
\begin{eqnarray}
 u_{t}+ 2\sqrt{1-u}(uv)_{x}&=&0, \label{102}\\
 v_{t}+(v^{2}\sqrt{1-u})_{x}&=&0. \label{103}
\end{eqnarray}
To derive this equation, we consider again the HFE (\ref{92}).  For the magnetic vector ${\bf A}$, we now consider the following parametrization  
\begin{eqnarray}
{\bf A} = 0.5\left(\phi+\bar{\phi}, i(\bar{\phi}-\phi), 2\sqrt{1-|\phi|^{2}}\right),  \label{104}
\end{eqnarray}
where $\phi$ is a complex function. Then the HFE  (\ref{92}) takes the form 
\begin{eqnarray}
i\phi_{t}+\sqrt{1-|\phi|^{2}}\phi_{xx}-\left(\sqrt{1-|\phi|^{2}}\right)_{xx}\phi=0.  \label{105}
\end{eqnarray}
Next we consider the transformation
\begin{eqnarray}
\phi=\sqrt{u}e^{i\epsilon^{-1}S}. 
\label{106}
\end{eqnarray}
Then the  dispersionless limit of the HFE   has the form
 \begin{eqnarray}
u_{t}+2\sqrt{1-u}(S_{x}u)_{x}&=&0, \label{107}\\
S_{t}+\sqrt{1-u}S_{x}^{2}&=&0. \label{108}
\end{eqnarray}
It is the M-LXXIV  equation written in terms of variables $u$ and $S$.  In terms of variables $u, v=S_{x}$, this equation turns to the form (\ref{102})-(\ref{103}). Note that the M-LXXIV  equation can be rewritten as
\begin{eqnarray}
 u_{\tau}+ 2(uv)_{x}&=&0, \label{109}\\
 v_{\tau}+2vv_{x}-\left[\frac{v^{2}u_{x}}{2(1-u)}\right]&=&0, \label{110}
\end{eqnarray}
where 
\begin{eqnarray}
 \frac{1}{\sqrt{1-u}}\frac{\partial}{\partial t}=\frac{\partial}{\partial \tau}. \label{111}
\end{eqnarray}

\subsection{M-LXXI equation}
In this section we consider the M-LXXI equation. This equation reads as
\begin{eqnarray}
 u_{t}+ 2(uv)_{x}- \frac{4uvu_{x}}{1+u}&=&0, \label{112}\\
 v_{t}+2vv_{x}- \left(\frac{2uv^{2}}{1+u}\right)_{x}+\frac{2j_{31}u}{1+u}&=&0, \label{113}
\end{eqnarray}
or
\begin{eqnarray}
 u_{t}+ 2(uS_{x})_{x}- \frac{4uu_{x}S_{x}}{1+u}&=&0, \label{114}\\
 S_{t}+S_{x}^{2}- \frac{2uS_{x}^{2}}{1+u}+2j_{31}x-2j_{31}\partial_{x}^{-1}\left(\frac{1}{1+u}\right)&=&0. \label{115}
\end{eqnarray}
Let us derive this equation. To do that consider the LLE
\begin{eqnarray}
 {\bf A}_{t}={\bf A}\wedge{\bf A}_{xx}+{\bf A}\wedge J{\bf A}, \label{116}
\end{eqnarray}
where $J=diag(j_{1},j_{2},j_{3})$ is an arbitrary constant diagonal matrix. Consider the transformation
\cite{Demskoi}:
\begin{eqnarray}
{\bf A} = \psi(1 - uv, i + iuv, u + v),  \label{117}
\end{eqnarray}
where 
\begin{eqnarray}
\psi=(u-v)^{-1}.  \label{118}
\end{eqnarray}
Then  the LLE takes the form \cite{Demskoi}:
\begin{eqnarray}
iu_{t}+u_{xx}-2\phi[u_{x}^{2}-P(u)]-\frac{1}{2}P^{'}(u)&=&0, \label{119}\\
iv_{t}-v_{xx}-2\phi[v_{x}^{2}-P(v)]+\frac{1}{2}P^{'}(v)&=&0. 
\label{120}
\end{eqnarray}
Here $P$ is a fourth degree polynomial of the form
\begin{eqnarray}
P(u) = \frac{1}{4}(j_{1}-j_{2})u^{4}-\frac{1}{2}(j_{1}+j_{2}-2j_{3})u^{2}+\frac{1}{4}(j_{1}-j_{2})= \frac{j_{12}}{4}u^{4}+\frac{j_{31}+j_{32}}{2}u^{2}+\frac{j_{12}}{4}, \label{121}
\end{eqnarray}
where $j_{ik}=j_{i}-j_{k}$. Note that the transformation (\ref{117})  coincides with the usual 
stereographic projection of the spin vector ${\bf A}$, 
if we set $v = -\bar{u}^{-1}$ \cite{Demskoi}.  For the variable $\omega=u$, the LLE takes the form
 \begin{eqnarray}
i\omega_{t}+\omega_{xx}-2\frac{\bar{\omega}[\omega_{x}^{2}-P(\omega)]}{1+|\omega|^{2}}-\frac{1}{2}P^{'}(\omega)&=&0. \label{122}
\end{eqnarray}
To find the dispersionless limit of this equation, we consider the transformation (\ref{97}). We also assume that $j_{12}=0$.  It is not difficult to verify that in this case the functions $u$ and $v$ obey the M-LXXI  equation (\ref{112})-(\ref{113}). Thus we have shown that the M-LXXI equation (\ref{112})-(\ref{113}) is one of the dispersionless limits of the LLE (\ref{116}).
\subsection{M-LXX equation}
In this section, our aim is to derive the M-LXX equation,  which has the following form
\begin{eqnarray}
 u_{t}+ 2\sqrt{1-u}(uv)_{x}&=&0, \label{123}\\
 v_{t}+(v^{2}\sqrt{1-u})_{x}+j_{31}(\sqrt{1-u})_{x}&=&0. \label{124}
\end{eqnarray}
To do that, again we return to  the LLE (\ref{116}).  For the magnetic vector ${\bf A}$, we now consider the parametrization (\ref{104}). Then the LLE (\ref{116}) takes the form 
\begin{eqnarray}
i\phi_{t}+\sqrt{1-|\phi|^{2}}\phi_{xx}-\left(\sqrt{1-|\phi|^{2}}\right)_{xx}\phi-0.5\sqrt{1-|\phi|^{2}}[(j_{31}+j_{32})\phi+j_{21}\bar{\phi}]=0.  \label{125}
\end{eqnarray}
For simplicity, let us consider the case when $j_{1}=j_{2}$. Then the equation (\ref{44}) takes the form
\begin{eqnarray}
i\phi_{t}+\sqrt{1-|\phi|^{2}}\phi_{xx}-\left(\sqrt{1-|\phi|^{2}}\right)_{xx}\phi-j_{31}\phi\sqrt{1-|\phi|^{2}}=0.  \label{126}
\end{eqnarray}
Next we consider the transformation (\ref{106}). 
Then the  dispersionless limit of the equation (\ref{126}) has the form
 \begin{eqnarray}
u_{t}+2\sqrt{1-u}(S_{x}u)_{x}&=&0, \label{127}\\
S_{t}+\sqrt{1-u}S_{x}^{2}+j_{31}\sqrt{1-u}&=&0. \label{128}
\end{eqnarray}
It is the M-LXX equation.  In terms of $u, v=S_{x}$, the M-LXX  equation (\ref{127})-(\ref{128}) turns to the form (\ref{123})-(\ref{124}). Note that the M-LXX equation can be rewritten as
\begin{eqnarray}
 u_{\tau}+ 2(uv)_{x}&=&0, \label{129}\\
 v_{\tau}+2vv_{x}-\left[\frac{(v^{2}+j_{31})u_{x}}{2(1-u)}\right]&=&0, \label{130}
\end{eqnarray}
where 
\begin{eqnarray}
 \frac{1}{\sqrt{1-u}}\frac{\partial}{\partial t}=\frac{\partial}{\partial \tau}. \label{131}
\end{eqnarray}

\section{Conclusions}

In this paper,  we have presented the pedagogical introduction to the theory of dispersionless equations. As examples of such equations, we consider well-known integrable dispersionless equations like dKdV, dNLS, dKP, dDS and so on.  In some cases, as examples we give Lax representations of some considered dispersionless equations. In particular, we give the  generalized  Benney equation  which 
describes some generalization of  Chaplygin gas equation. We have considered  the
Lax representation for this Benney equation. 
All these equations are likely to be integrable since they follow from
a Lax description. However, we have not studied these systems in more
detail.

\section{Acknowledgements}
This work was supported in part by the Ministry of Edication  and Science of Kazakhstan under
grant 0118RK00935  as well as by grant  0118RK00693.

\end{document}